\documentclass{bmcart}
\usepackage{graphics,epsfig,placeins,subfigure,wrapfig}
\usepackage{amsthm,amsmath}
\RequirePackage{natbib}
\RequirePackage{hyperref}
\usepackage[utf8]{inputenc} 
\usepackage{url}

\startlocaldefs
\endlocaldefs

\begin{document}

\begin{frontmatter}

\begin{fmbox}
\dochead{Survey}


\title{Convergence of Artificial Intelligence and \\ High Performance Computing on \\ NSF-supported Cyberinfrastructure}


\author[
   addressref={aff1,aff1.5,aff2,aff3},                   
   corref={aff1,aff1.5,aff2,aff3,aff4,aff5,aff6,aff7,aff8},                       
   noteref={n1},                        
   email={elihu@illinois.edu}   
]{\inits{EA}\fnm{E. A.} \snm{Huerta}}
\author[
   addressref={aff1,aff1.5,aff3},
   noteref={n1},  
]{\inits{A}\fnm{Asad} \snm{Khan}}
\author[
   addressref={aff8},
   noteref={n1},  
]{\inits{ED}\fnm{Edward} \snm{Davis}}
\author[
   addressref={aff1},
]{\inits{CB}\fnm{Colleen} \snm{Bushell}}
\author[
   addressref={aff1,aff1.5,aff4},
]{\inits{WG}\fnm{William D.} \snm{Gropp}}
\author[
   addressref={aff1,aff4,aff5,aff6},
]{\inits{DSK}\fnm{Daniel S.} \snm{Katz}}
\author[
   addressref={aff1,aff1.5,aff4,aff5},
   noteref={n1},  
]{\inits{VK}\fnm{Volodymyr} \snm{Kindratenko}}
\author[
   addressref={aff1,aff1.5,aff7},
]{\inits{SK}\fnm{Seid} \snm{Koric}}
\author[
   addressref={aff1,aff4},
]{\inits{WTCK}\fnm{William T. C.} \snm{Kramer}}
\author[
   addressref={aff1,aff1.5},
]{\inits{BM}\fnm{Brendan} \snm{McGinty}}
\author[
   addressref={aff1},
]{\inits{KM}\fnm{Kenton} \snm{McHenry}}
\author[
   addressref={aff1,aff1.5},
]{\inits{AS}\fnm{Aaron} \snm{Saxton}}

\address[id=aff1]{
  \orgname{National Center for Supercomputing Applications, University of Illinois at Urbana-Champaign}, 
  \street{Urbana},                     %
  \postcode{61801}                                
  \city{Illinois},                              
  \cny{USA}                                    
}

\address[id=aff1.5]{
  \orgname{NCSA Center for Artificial Intelligence Innovation, University of Illinois at Urbana-Champaign}, 
  \street{Urbana},                     %
  \postcode{61801}                                
  \city{Illinois},                              
  \cny{USA}                                    
}

\address[id=aff2]{%
  \orgname{Department of Astronomy, University of Illinois at Urbana-Champaign},
  \street{Urbana},
  \postcode{61801}
  \city{Illinois},
  \cny{USA}
}

\address[id=aff3]{%
  \orgname{Department of Physics, University of Illinois at Urbana-Champaign},
  \street{Urbana},
  \postcode{61801}
  \city{Illinois},
  \cny{USA}
}

\address[id=aff4]{%
  \orgname{Department of Computer Science, University of Illinois at Urbana-Champaign},
  \street{Urbana},
  \postcode{61801}
  \city{Illinois},
  \cny{USA}
}

\address[id=aff5]{%
  \orgname{Department of Electrical and Computer Engineering, University of Illinois at Urbana-Champaign},
  \street{Urbana},
  \postcode{61801}
  \city{Illinois},
  \cny{USA}
}

\address[id=aff6]{%
  \orgname{School of Information Sciences, University of Illinois at Urbana-Champaign},
  \street{Urbana},
  \postcode{61801}
  \city{Illinois},
  \cny{USA}
}

\address[id=aff7]{%
  \orgname{Department of Mechanical Science and Engineering, University of Illinois at Urbana-Champaign},
  \street{Urbana},
  \postcode{61801}
  \city{Illinois},
  \cny{USA}
}

\address[id=aff8]{%
  \orgname{University of Queensland},
  \street{St Lucia},
  \postcode{QLD 4072}
  \cny{Australia}
}


\begin{artnotes}
\note[id=n1]{Equal contributor} 
\end{artnotes}

\end{fmbox}


\begin{abstractbox}

\begin{abstract} 
Significant investments to upgrade and construct large-scale scientific facilities demand commensurate investments in 
R\&D to design algorithms and computing approaches to enable scientific and engineering breakthroughs in the big data 
era. Innovative Artificial Intelligence (AI) applications have powered transformational solutions for big data challenges in 
industry and technology that now drive a multi-billion dollar industry, and which play an ever increasing role shaping human 
social patterns. As AI continues to evolve into a computing paradigm endowed with statistical and mathematical rigor, it has  
become apparent that single-GPU solutions for training, validation, and testing are no longer sufficient for computational grand challenges brought about by scientific facilities that produce data at a rate and volume that outstrip the computing capabilities of available cyberinfrastructure platforms. This realization has been driving the confluence of AI and high performance computing (HPC) to reduce time-to-insight, and to enable a systematic study of domain-inspired AI architectures and optimization schemes to enable data-driven discovery. In this article we present a summary of recent developments in this field, and describe specific advances that authors in this article are spearheading to accelerate and streamline the use of HPC platforms to design and apply accelerated AI algorithms in academia and industry.

%
%
\end{abstract}




\end{abstractbox}
%

\end{frontmatter}



\section*{Introduction}

The big data revolution disrupted the digital and computing landscape in the early 2010s~\cite{HPCUSE}. Data torrents produced by corporations such as Google, Amazon, Facebook and YouTube, among others, presented a unique opportunity for innovation. Traditional signal processing tools and computing methodologies were inadequate to turn these big-data challenges into technological breakthroughs. A radical rethinking was urgently needed~\cite{NAP25199,goodfellow_dl}. 

Large Scale Visual Recognition Challenges~\cite{ILSVRC15} set the scene for the ongoing digital revolution. The quest for novel pattern recognition algorithms~\cite{Lecun:1998, Lecun:2015Nature,LeCun:bp} that sift through large, high-quality data sets eventually led to a disruptive combination of deep learning and graphics processing units (GPUs) that enabled a rapid succession of advances in computer vision, speech recognition, natural language processing, and robotics, to mention a few~\cite{resnet_human,goodfellow_dl}. These developments are currently powering the renaissance of AI, which is the engine of a multi-billion dollar industry. 

Within just a few years, the curation of high-quality data sets, e.g., \texttt{ImageNet}~\cite{imagenet_cvpr09}; GPU-accelerated computing~\cite{Krizhevsky:2015}; open source software platforms---\texttt{TensorFlow}~\cite{tensorflow2015-whitepaper}, \texttt{PyTorch}~\cite{NEURIPS2019_9015} among others---to design, train, validate and test AI models; improved AI architectures and novel techniques~\cite{pi_dl,asad:2020K} to enhance the performance of deep neural networks, such as robust optimizers~\cite{kingma2014adam} and regularization techniques~\cite{kukaka2017regularization}, led to the rapid development of AI tools that significantly outperform other signal processing tools on many tasks~\cite{SCHMIDHUBER201585,Sejnowski201907373}. Data-driven discovery is now also informing and stirring the design of exascale cyberinfrastructure, in which high performance computing (HPC) and data have become a single entity, namely HPCD~\cite{NAP25199,NAP21886}.


\section*{Convergence of AI and HPC}

The convergence of AI and HPC is being pursued in earnest across the HPC ecosystem. Recent accomplishments of this program have been reported in plasma physics~\cite{Svyatkovskiy:GPU}, cosmology~\cite{asad:2018K}, gravitational wave astrophysics~\cite{Shen:2019DLScale}, high energy physics~\cite{hep_kyle}, multi-messenger astrophysics~\cite{Huerta:2019rtg}, materials science~\cite{ward_blaiszik_foster_assary_narayanan_curtiss_2019}, data management of unstructured datasets~\cite{Marini:2018,Padhy:2015}, and genetic data~\cite{Blatti642124}, among others. 
 
These achievements share a common thread, namely, the algorithms developed to accelerate the training of AI models in HPC platforms have a strong experimental component. To date, there is no rigorous framework to constrain the ideal set of hyper-parameters that ensures rapid convergence and optimal performance of AI models as the number of GPU nodes is increased to accelerate the training stage. Furthermore, it is customary that distributed training algorithms in HPC platforms are benchmarked using idealized neural network models and datasets, e.g., training a \texttt{ResNet} model~\cite{he2015deep} using the \texttt{ImageNet} dataset~\cite{imagenet_cvpr09}. While this approach provides some guidance about the optimal performance of HPC platforms for deep learning research, it does not impart any insights regarding the actual performance of these facilities when using domain-inspired AI architectures and optimization schemes to do data driven discovery in the context of realistic datasets, which are noisy, incomplete, and heterogenous---vastly different from the \texttt{ImageNet} dataset. 

In view of these considerations, some key developments are needed to maximize the potential of 
AI for data-driven discovery: (i) the development of a rigorous mathematical framework to make informed choices of domain inspired AI architectures and optimization schemes; (ii) the creation of an interdisciplinary effort that brings together domain, information science, AI, data and software experts to inform the collection and 
curation of experimental and simulation datasets; (iii) the identification of connections between AI data and models, which will 
facilitate the production of commodity software that may be seamlessly applicable to disparate fields that 
share common data and computing data challenges; and (iv) the deployment of AI models and data on open source platforms, such as the Data and Learning Hub for Science~\cite{dlhub,blaiszik_foster_2019}. These activities will accelerate the adoption of reproducible and robust AI tools as commodity software across disciplines.

There are several dedicated efforts in the literature to address these timely and relevant 
challenges, 
see e.g.~\cite{8638041,8030298,Frankle2019TheLT}. In the US, the National Science Foundation (NSF) and 
the Department of Energy (DOE) are 
spearheading multi-million dollar programs to spearhead the construction of the next generation 
of HPC platforms to address computational grand challenges at the exascale, and on 
R\&D to accelerate the design, deployment and adoption of 
innovative AI applications for data-driven discovery and science and engineering, and to translate these 
innovations into tangible societal benefits, business and industry. The funding of new HPC 
platforms for innovative AI 
research such as Bridges-2, Delta, and Neocortex will provide transformative capabilities by 
introducing new hardware for AI research~\cite{new_systems,bridges-2}. The Frontier, Aurora 
and El Capitan exascale systems will combine simulation, data science, and machine learning 
to revolutionize how supercomputers are used for scientific discovery and innovation.

In terms of R\&D, DOE has launched an initiative to make AI models and data that adhere to FAIR data principles (Findable, Accessible, Interoperable, and Reusable). The goal of this program is to set a standard for the production of data that may be reusable both by researchers and machines, with little human intervention. It is expected that this approach will enable researchers to gain new insights on how AI models abstract knowledge from data, and to quantify how domain-inspired optimization schemes guide AI to convergence to the right answer in controlled experiments, while also enabling intuitive AI discovery that is beyond the reach of existing theories that do not fully capture complex phenomena, such as turbulence~\cite{Rosofsky:2019}. This program will maximize the use of exascale HPCD platforms, accelerating the development of AI.

While it is customary to quantify the performance of HPC platforms for distributed training at scale using idealized datasets and vanilla AI models, i.e., \texttt{ResNet-50} trained with the \texttt{ImageNet} dataset, it is also 
important to assess the performance of advanced cyberinfrastructure facilities to train more complex, domain-inspired AI models with realistic, experimental datasets. To provide a broad perspective on the state-of-the-art for different domains, we present results for a number of studies that we have conducted on NSF and DOE HPC platforms. The AI models we consider are tailored for image recognition, classification and regression analyses of telescope image datasets, and time-series data that describe the collision of black holes. To showcase the use of these models and datasets, we have used two NSF funded HPC platforms, namely, the Hardware-Accelerated Learning (HAL) cluster~\cite{halcluster} at the National Center for Supercomputing Applications (NCSA), and the Bridges-AI system~\cite{bridgesai} that is part of the Extreme Science and Engineering Discovery Environment (XSEDE) at the Pittsburgh Supercomputing Center (PSC); and the DOE-funded Summit supercomputer at Oak Ridge National Laboratory~\cite{summit_hpc}.

\noindent \textbf{HPC Platforms} The HAL cluster has 64 NVIDIA V100 GPUs distributed evenly across 16 nodes, and connected by NVLink 2.0~\cite{halcluster} inside the nodes and EDR InfiniBand across the nodes. In Bridges-AI~\cite{bridgesai} we have used the 9 HPE Apollo 6500 servers, each with 8 NVIDIA Tesla V100 GPUs with 16 GB of GPU memory each, connected by NVLink 2.0. 

\begin{figure}[h!]
	\centerline{
	\includegraphics[width=\textwidth]{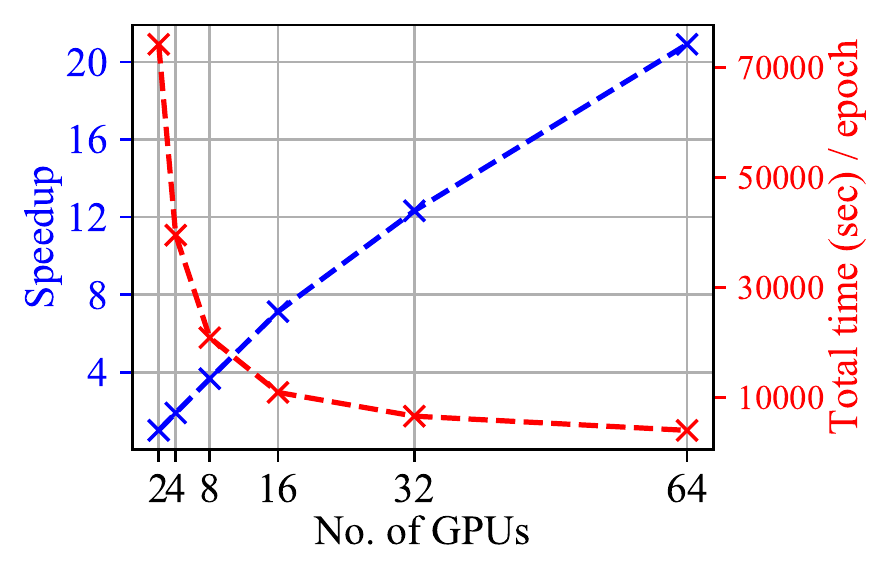}
	}
\caption{\csentence{\texttt{ImageNet ResNet-50} training} Global throughput (images/sec) and speed up obtained by scaling the training of \texttt{ResNet-50} using the \texttt{ImageNet} dataset. The training stage is reduced to just over 1 hour, achieving 93\% accuracy, using the entire HAL cluster. }
\label{fig:fig_resnet}
 	\end{figure}

\begin{figure}[h!]
  	\centerline{
	\includegraphics[width=\textwidth]{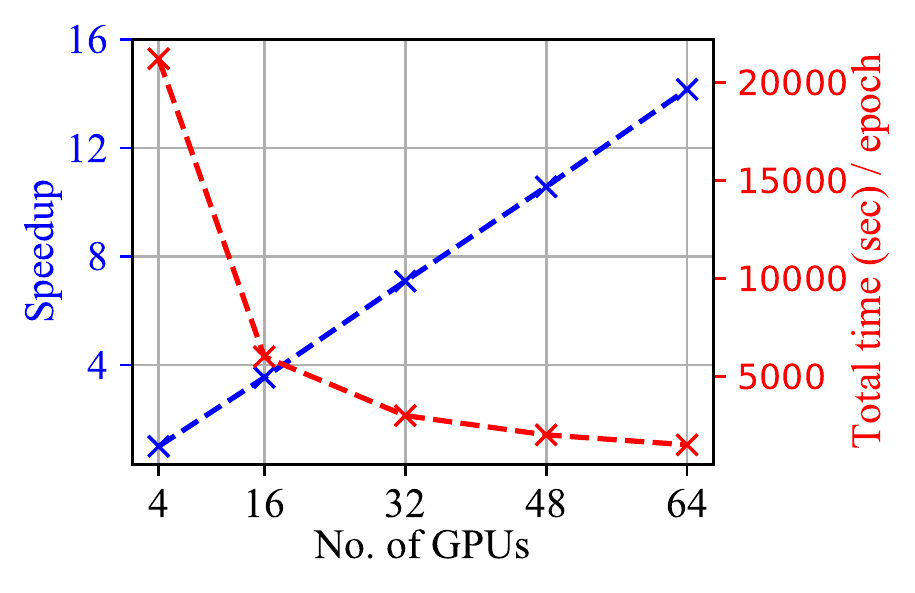}
	}
  \caption{\csentence{Gravitational Wave Astrophysics with the HAL Deep Learning Cluster}
      The training stage of a deep learning model, used to infer how rapidly two colliding black holes rotate, is reduced from 1 month---using a single V100 GPU---to 12.4 hours using the entire HAL deep learning cluster at the National Center for Supercomputing Applications.}
      \label{fig:fig_1}
      \end{figure}

\begin{figure}[h!]
        	\centerline{
	\includegraphics[width=\textwidth]{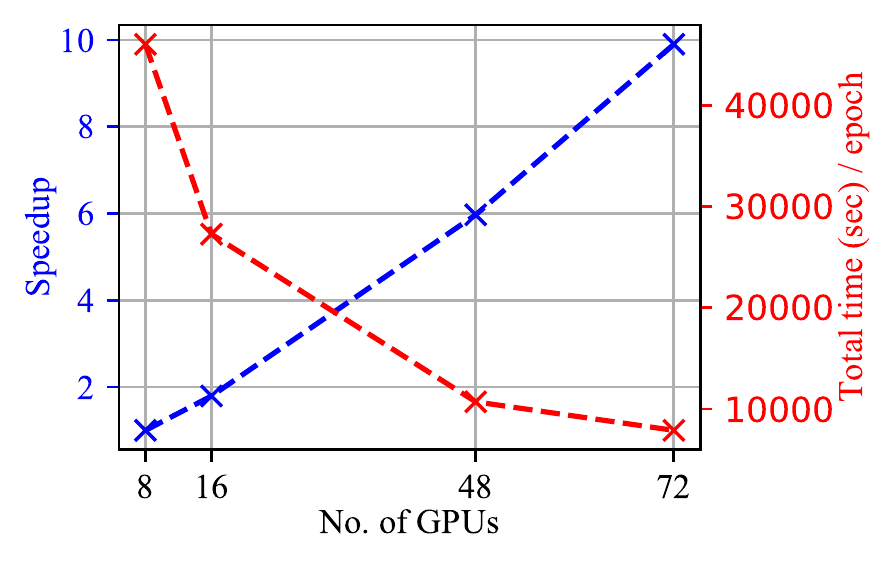}
	}
  \caption{\csentence{Gravitational Wave Astrophysics with the XSEDE Bridges-AI Cluster}
      As Figure~\ref{fig:fig_1}, but now using the entire Bridges-AI cluster at the Pittsburgh Supercomputing Center. In this case, we reduce the training stage to 38 hours using 72 V100 GPUs.}
      \label{fig:fig_2}
      \end{figure}

\begin{figure}[h!]
          	\centerline{
	\includegraphics[width=\textwidth]{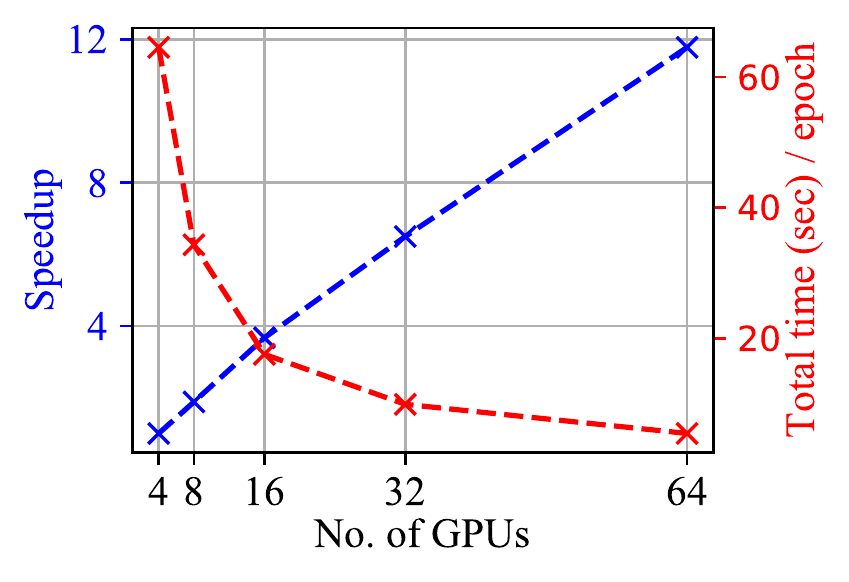}
	}
\caption{\csentence{Cosmology with the HAL Deep Learning Cluster}
      The training stage of a deep learning model, used to morphologically classify galaxies between spiral and elliptical classes,  is reduced from 2.1 hours---using a single V100 GPU---to just 2.7 minutes using the entire HAL deep learning cluster.}
      \label{fig:fig_3}
      \end{figure}

\noindent \textbf{AI Models and Datasets} We have used three different AI models: (i) \texttt{ResNet-50}; (ii) an AI model to characterize the signal manifold of binary black hole mergers, trained with time-series signals that describe gravitational wave signals~\cite{asad:2020K} (\texttt{AI-GW}); and (iii) an AI model that classifies galaxy images collected by the Sloan Digital Sky Survey (SDSS)~\cite{SDSS_summary}, and automatically labels galaxy images collected by the Dark Energy Survey (DES)~\cite{asad:2018K} (\texttt{AI-DES}). Our results of these analyses indicate: 

\begin{itemize}
\item  Figure~\ref{fig:fig_resnet} shows that \texttt{ResNet-50} with \texttt{ImageNet} is trained within 41 hours using 1 V100 GPU in HAL. The training is reduced to just over 1 hour, achieving 93\% accuracy, using 64 V100 GPUs in HAL.
\item Figure~\ref{fig:fig_1} shows that \texttt{AI-GW} is fully trained, achieving state-of-the-art accuracy, within 754 hrs using a single V100 GPU in HAL. When scaled to 64 V100 GPUs, the training is reduced to 12.4 hours. 
\item Figure~\ref{fig:fig_2} shows that \texttt{AI-GW} is fully trained, achieving state-of-the-art accuracy, within 38 hours using 72 V100 GPUs in Bridges-AI.
\item Figure~\ref{fig:fig_3} shows that \texttt{AI-DES} is trained within 2.1 hrs using a single V100 GPU in HAL. The training is reduced to 2.7 minutes using 64 V100 GPUs in HAL. 
\end{itemize}

\noindent These examples clearly underscore the importance of coupling AI with HPC: (i) it significantly speeds up the training stage, enabling the exploration of domain-inspired architectures and optimization schemes, which are critical for the design of rigorous, trustworthy and interpretable AI solutions; (ii) it enables the use of larger training data sets to boost the accuracy and reliability of AI models while keeping the training stage at a minimum.


\section*{Software and Hardware Challenges}

While open source software platforms have played a key role in the swift evolution of AI, they present a number of challenges when used in HPC platforms. This is because open source software platforms such as \texttt{TensorFlow}~\cite{tensorflow2015-whitepaper} and \texttt{PyTorch}~\cite{NEURIPS2019_9015} are updated at a much faster pace than libraries deployed cluster-wide on HPC platforms. For instance, in typical HPC platforms, software updates customarily take place twice per year~\cite{ibm_updates,ibm_updates_1}. In the case of open source AI APIs, releases happen much more often, as can be seen in the official release timeline of \texttt{TensorFlow}~\cite{tf_release}. Furthermore, producing AI models usually requires a unique set of package dependencies. Therefore, the traditional use of modules has limited effectiveness since software dependencies change between projects and sometimes evolve even during a single project. Common solutions to give users more fine-grained control over software environments include containerization, e.g., \texttt{Singularity}~\cite{kurtzer:2016} or \texttt{Kubernetes}~\cite{kuber}, and virtual environments such as \texttt{Anaconda}~\cite{anaconda} that is provided in HPC platforms such as Bridges, Bridges-AI, Summit, and HAL. GPUs play a key role in the renaissance of AI because they have unique features to accelerate applications, e.g., they have many cores, provide high throughput, they are good for parallel processing and can perform thousands of operations at once. While these features are particularly relevant for image recognition analysis, gaming and graphics, GPUs are now used extensively in other areas, i.e., autonomous driving and robotics. In the context of HPC and AI, our studies indicate that 5 nodes (each node has 64 Intel KNL 7230 compute cores) in Theta are equivalent to a single V100 GPU. Thus, given how involved it is to optimally scale the training of AI models in HPC platforms, it is apparent the advantage provided by GPU-based HPC platforms for AI research.

We provide below a number of recommendations to streamline the use of HPC resources for AI research:

\begin{enumerate}
\item Provide up-to-date documentation and tutorials to set up containers and virtual environments, and adequate help desk support to enable smooth, fast-paced project life-cycles. 
\item Maintain a versatile, up-to-date base container image, and base virtual environment that users can easily clone and modify for their specific needs.
\item Distributed training software stacks such as \texttt{TensorFlow} depend on distributed training software stacks, e.g., \texttt{Horovod}~\cite{2018arXiv180205799S}, which in turn depend on system architecture and specific versions of \texttt{MPI} installed by system and service managers. It is important to have clear up-to-date documentation on system architecture and \texttt{MPI} versions installed, and clear instructions on how to install/update distributed training software packages like \texttt{Horovod} into the user's container/virtual environment.
\end{enumerate}

\noindent In addition to these considerations, the AI model architecture, dataset, and training optimizer prevent a seamless use of distributed training. Stochastic gradient descent (SGD)~\cite{sgd_paper} and its variants are the workhorse optimizer for AI training. The common way to parallelize training is to use ``mini-batches'' with SGD. In principle, a larger mini-batch may naively utilize more GPUs (or CPUs). Training time to solution will often scale linearly with small batch size. Figures~\ref{fig:fig_1} and~\ref{fig:fig_3} show good generalization at 64 GPUs, which amounts to a global batch size of 128 samples.  However, it is known that as data sets and number of features grow, naively scaling number of GPUs, and subsequently batch size, will often take more epochs to achieve an acceptable validation error. The state-of-the art in AI training at scale was reported in~\cite{Jia:2018}. Therein, \texttt{ResNet} was trained using a batch size of 64k samples, run across 2048 Tesla P40s. While achieving this level of scaling required a lot of experimental work, this benchmark, and others~\cite{You:2018}, indicate that scaling AI models to larger data and feature sets is indeed possible. However, it requires a considerable amount of human effort to tune the model and training pipeline. A mixture of fast human model development cycle mixed with automated hyper-parameter tuning is a candidate solution to tackle this problem. 

\begin{figure}[h!]
  	\centerline{
	\includegraphics[width=\textwidth]{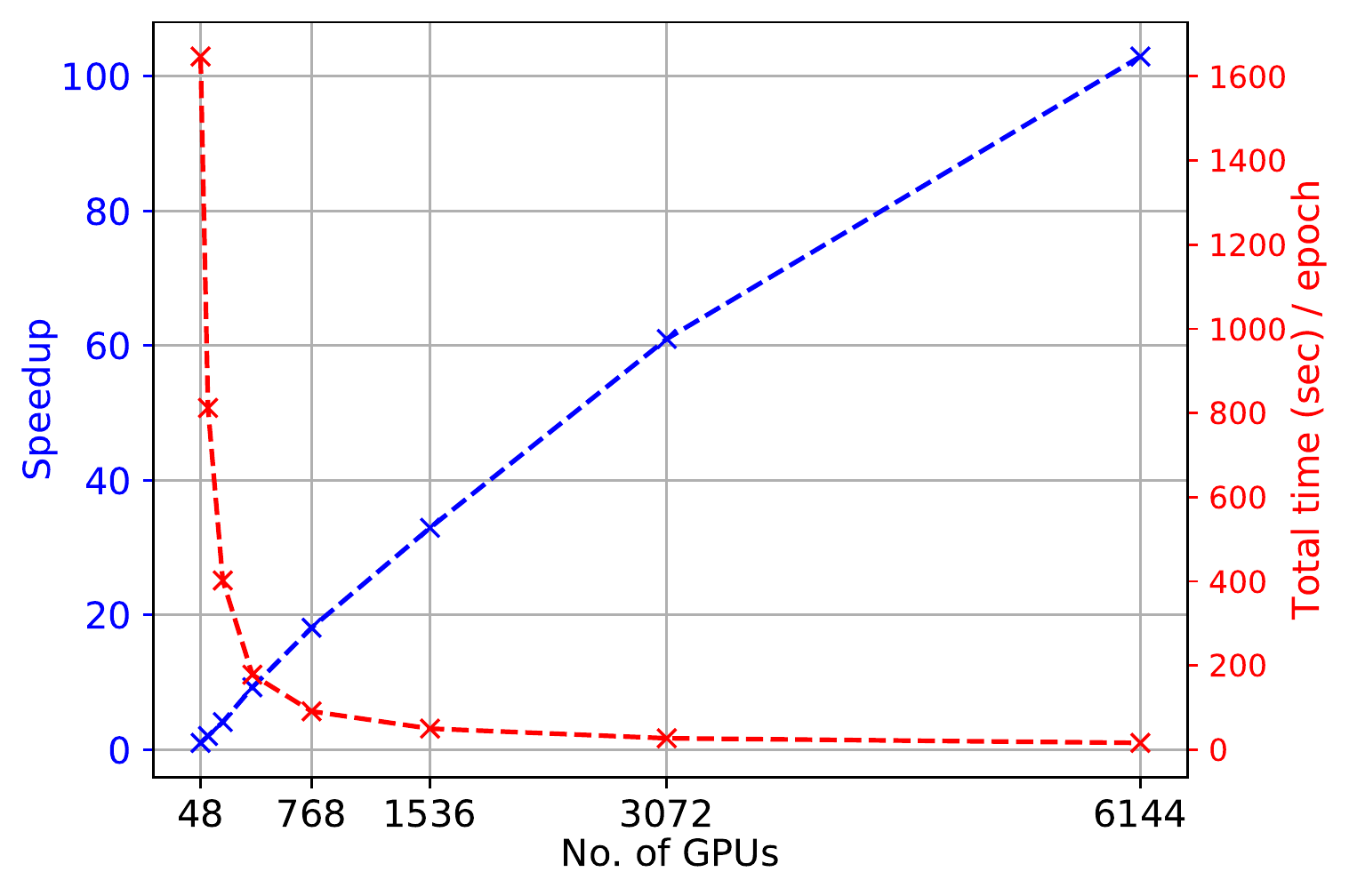}
	}
        \caption{\csentence{Gravitational Wave Astrophysics with Summit}
      As Figure~\ref{fig:fig_1}, but now using 1,536 V100 GPUs in the Summit supercomputer at Oak Ridge National Laboratory. At this scale, the model is trained in 1,2 hours.}
      \label{fig:fig_4}
      \end{figure}

We have explored whether the methods we have used in the context of HAL and Bridges-AI may work in other HPC platforms optimized for AI research. In Figure~\ref{fig:fig_4} we show that our distributed training algorithms exhibit strong scaling up to 1024 nodes (6144 V100 GPUs) in the Summit supercomputer at Oak Ridge National Lab. The scaling efficiency, i.e., how long it takes to cycle through all of the data once, also known as Total time / epoch (see y-axis label on the right of Figure~\ref{fig:fig_4}) can be affected by many factors, e.g., I/O speed, communication, etc., and achieving good efficiency and strong scaling, as shown in this Figure, indicates that we have dealt with properly with these factors. 

Furthermore, Figure~\ref{fig:fig_4} shows that using 256 nodes (1,536 V100 GPUs) in the Summit supercomputer we are able to fully train a physics-inspired version of the \texttt{WaveNet} model with time-series data that describes numerical solutions to Einstein's equations that model black hole collisions, attaining state-of-the-art accuracy, within just 1.2 hours. In other words, we can generalize the methods deployed and tested on NSF-funded cyberinfrastructure to HPC platforms that have different scale, hardware and software.

\noindent \textbf{Open Challenges} A number of challenges remain towards an optimal exploitation of AI and extreme scale computing. For instance, it is recognized that some experimental datasets are not in a suitable format to fully exploit data-driven discovery. To address this pressing issue, DOE has made significant investments to make AI models and data FAIR~\cite{doe_fair}. Another challenge concerns the design of AI models whose architecture and optimization schemes incorporate domain knowledge, enabling AI models to converge faster while also enabling intuitive, serendipitous discovery that may not be encapsulated by approximate descriptions of complex phenomena~\cite{2017NatPhVMat,Rosofsky:2019}. It is also essential to develop a rigorous approach to maximize the use of HPC platforms for distributed training. This requires a systematic approach to select an optimal set of hyperparameters that enables faster convergence, and creative methods to use less training data to achieve state-of-the-art performance. NSF has also funded several institutes to advance the state-of-the-art in AI, seeking new modes of data-driven discovery in science and engineering. These investments aim to sustain, broaden and accelerate recent breakthroughs in science, technology and industry driven by AI applications~\cite{ai_nsf}. As these projects evolve and mature, it will be essential to facilitate cross-pollination of expertise, avoiding duplication and empowering new AI practitioners to access AI scientific software that is open source, interpretable, reproducible and trustworthy.

\section*{Cloud Computing and HPC}
Cloud computing and containerization became popular for developing customer facing web apps. It allowed a DevOps team---i.e., the team that develops scientific software and manages ongoing operations of a data center---to keep strict control of the customer facing software, while new features and bug fixes were designed, developed, and tested in an environment that ``looked the same'' as a live one. Depending on the business cycle, companies could dynamically scale their infrastructure with virtually no overhead of purchasing hardware, and then relinquish it when it was no longer needed.

HPC would do well to adopt a DevOps cycle like the ones seen in startup culture. However HPC has some unique challenges that make this difficult. 1) Data storage separated from compute in the form of a shared file system and an instance on maintaining a traditional tree like file system. Cloud computing delivers a unit of compute and storage in tandem as a single instance and isolates distinct resources. A developer using cloud resources treats a compute instance as only the host for their code and must explicitly choose how to move large volumes of data on and off. This is usually done by allocating a specialized cloud instance of a data store, e.g., SQL databases. Improved cloud solutions provide \texttt{Kubernetes} (and other cluster manager) recipes to allocate a skeleton of these resources, but it is still up to the developers to choose exactly how data are moved between the resources and to code the specific functions of their app. 2) HPC is a shared resource. That is, many users with different projects see the same file system and compute resource. Each developer must wait their turn to see their code run. In cloud computing, a resource belongs and is billed to the developer on demand. When the resource is released, all of its state-full properties get reset. 3) HPC is very concerned with the compute resources interconnect. To have high bandwidth and low latency between cloud compute instances, one pays a premium.

In the case of distributed training, one needs to ascertain whether the cloud or HPC platforms provide an adequate solution. On-demand, high throughput or cloudbursting of single-node applications are ideally suited for the cloud. For instance, in the case of genetic data analysis, the \texttt{KnowEng} platform~\cite{Blatti642124} is implemented as a web application where the compute cluster is managed by \texttt{Kubernetes}, and provides an example of a workflow that can be expanded to include methods for intuitively managing library compatibility and cloud bursting. This cloud-based solution includes: (1) the ability to access disparate data; (2) set parameters for complex AI experiments effortlessly; (3) deploy computation in a cloud environment; (4) engage with sophisticated visualization tools to evaluate data and study results; and (5) save results and access parameter settings of prior runs. 

However, large distributed training workloads that run for many hours or days will continue to excel on a high-end HPC environment. For instance, the typical utilization of the HAL cluster at NCSA tends to be well above 70\%. Given that the cost of a single V100 GPU node on AWS (p3.2xlarge instance~\cite{aws_money}) is \$3.06 per hour, HAL provides over \$141,000 in comparable cloud compute resources every month. This is far higher than the amortized cost of the HAL cluster and its support. A top-tier system like Blue Waters, where a node hour is charged at \$0.60, 4,228 K20 GPUs might have a cloud cost of \$2-3M per month.

\section*{Industry Applications}
The confluence of AI and HPC is a booming enterprise in the private sector. NCSA is spearheading its application to support industry partners from the agriculture, healthcare, energy, and financial, sectors to stay competitive on the global market by analyzing bigger and more complex data to uncover hidden patterns, reveal market and cash flow trends, and identify customer preferences~\cite{ncsaindustry}. The confluence of modeling, simulation and AI is another area of growing interest among manufacturing and life science partners, promising to significantly accelerate many extremely difficult and computationally expensive methods and workflows in model-based design and analysis~\cite{Abueidda:2020,Shirui:2020,Rosofsky:2019}.

Academic innovation in AI pursues ideas that are exciting and productive, though they may not have immediate, tangible benefits. While academic scholarship is curiosity driven research, innovative AI applications in industry have as a goal to address computational grand challenges at an accelerated pace, and to apply at scale new solutions to profit from them. In brief, while academia and industry pursue distinct goals, it is essential that both spheres of activity maintain a close-knit collaboration~\cite{aca_ind}. This is a critical endeavor because breakthroughs in industry and technology over the last decade were enabled by basic AI applications. As industrial applications reach new frontiers and computational grand challenges arise, it will be essential to continue leveraging AI innovation, and explore ways to translate it into tangible solutions that may be deployed at scale to produce societal and business benefits. In summary, the training of future AI practitioners demands an interdisciplinary approach that includes a clear vision of industry needs. This approach will ensure that academic AI innovation is readily incorporated and applied, creating a sustainable paradigm that opens up diverse lines of funding for AI researchers. 

\section*{Conclusion}
The convergence of AI and HPC provides the means to address big data challenges in science, engineering and industry, and enables the creation of disruptive approaches for data-driven discovery and innovation. Realizing these goals demands a concerted effort between AI practitioners, HPC and domain experts. 

As AI and HPC continue to transform an ever increasing number of disciplines at an accelerated pace, we can only image what the future holds once AI is powered with a rigorous mathematical framework. In that scenario, it will be possible to optimally use oversubscribed HPC platforms, and create intuitive AI solutions that will lead to transformational scientific discoveries, and disruptive solutions in industry and technology

Finally, to contribute to the use of realistic datasets to benchmark HPC platforms, we release two neural network models, along with datasets, that we used to produce Figures~\ref{fig:fig_1}--\ref{fig:fig_4}. As the NSF and other funding agencies continue to deploy faster and more powerful HPC platforms for AI research, it is urgent that we provide guidelines to maximize the use of these resources, and continue training new talent that will catalyze the adoption and best AI practices. This approach was critical in the past to enable the adoption of HPC by industry, and will play a more significant role in the future given the eagerness with which industry is adopting AI solutions.

\section*{List of Abbreviations}
AI: artificial intelligence; GPUs: graphics processing units; HPC: high performance computing; 
R\&D: research and development; NCSA: National Center for Supercomputing Applications; 
NSF: National Science Foundation; HAL: Hardware-Accelerated Learning; 
XSEDE: Extreme Science and Engineering Discovery Environment; PSC: Pittsburgh Supercomputing Center; 
GW: gravitational wave; SDSS: Sloan Digital Sky Survey; DES: Dark Energy Survey.


\begin{backmatter}

\section*{Ethics approval and consent to participate}
Not applicable

\section*{Consent for publication}
The authors approve the publication of this manuscript

\section*{Availability of data and materials}
The neural network models and data used to study characterize black hole mergers, and to classify galaxy images, are are readily available at the Deep Learning Hub (DLHub)~\cite{dlhub,blaiszik_foster_2019} hosted by Argonne National Laboratory (ANL)~\cite{dlhubmodel1,dlhubmodel2}.

\section*{Competing interests}
  The authors declare that they have no competing interests.

\section*{Funding}
EAH, AK, DSK, and VK gratefully acknowledge National Science Foundation (NSF) award OAC-1931561. EAH and VK also acknowledge NSF award OAC-1934757. This work utilized XSEDE resources through the NSF award TG-PHY160053, and the NSF's Major Research Instrumentation program, award OAC-1725729, as well as the University of Illinois at Urbana-Champaign. This research used resources of the Oak Ridge Leadership Computing Facility, which is a DOE Office of Science User Facility supported under Contract DE-AC05-00OR22725.

\section*{Author's contributions}
EAH led and coordinated the writing of this article. ED, EAH, AK and VK produced results for Figures~\ref{fig:fig_1}---\ref{fig:fig_resnet}. EAH and AK produced results for Figures~\ref{fig:fig_2} and~\ref{fig:fig_4}. All authors contributed to developing the ideas, and writing and reviewing this manuscript. 

\section*{Acknowledgements}
We thank Nicholas A. Nystrom, Paola Buitrago and Julian Uran for their support using Bridges-AI; and  Arjun Shankar, Tom Gibbs, Junqi Yin, and Jeff Larking for their support and guidance using the Summit supercomputer. We also thank Ben Blaiszik, Ryan Chard and Logan Ward for their support deploying our neural network models and testing datasets at the Data and Learning Hub for Science hosted by Argonne National Lab.  

\bibliographystyle{unsrt}
\bibliography{survey_paper.bib}      





\vspace{5mm} 




\section*{Additional Files}
  \subsection*{Additional file 1 --- Neural Network Model and Data to Characterize Black Hole Mergers}
    A fully trained neural network model, and time-series testing dataset, are readily available at the Deep Learning Hub (DLHub)~\cite{dlhub,blaiszik_foster_2019} hosted by Argonne National Laboratory (ANL)~\cite{dlhubmodel1}.

  \subsection*{Additional file 2 --- Neural Network Model and Data to Characterize Galaxy Images}
    A fully trained neural network model, and image testing dataset, are readily available at the DLHub hosted by ANL~\cite{dlhubmodel2}.

  \subsection*{Additional file 3 --- \texttt{ResNet-50} trained with \texttt{ImageNet}}
    All the code used for this work is open source at~\cite{vlad_HAL,vlad_hal_paper}.

\end{backmatter}
\end{document}